\documentclass[12pt,twoside]{article}
\usepackage{latexsym}
\usepackage{amsbsy}
\usepackage[dvips]{graphicx}
\usepackage{afterpage}
\newcommand{\bbox}[1]{\mbox{\boldmath $#1$}}
\def\vec#1{{\bbox #1}}

\let\oldsection=\section
\renewcommand{\section}[1]{\oldsection{\large #1}}
\begin{document}
\begin{titlepage}    
    \vspace{-1cm}
    \begin{flushright}
ISN-01-114\\ Lycen-2001-75\\ physics/0111111
\end{flushright}
\vfill
\begin{center}
{\bf\Large Stability of three-and four-charge systems$^{*}$ }
\vfill
{\large 
Jean-Marc Richard}$^\dagger$
\\[1cm]
{\small Institut de Physique Nucl\'eaire}\\[-.15cm]
{\small Universit\'e Claude Bernard--CNRS-IN2P3}\\[-.15cm]
{\small 4, rue Enrico Fermi, F--69622 Villeurbanne cedex, France}\\[.15cm]
{\small and}\\ [.15cm]
{\small Institut des Sciences Nucl\'eaires}$^\ddagger$\\[-.15cm]
{\small Universit\'e Joseph Fourier--CNRS-IN2P3}\\ [-.15cm]
{\small 53, avenue des Martyrs, F--38026  Grenoble cedex, France}
\end{center}
\vfill
\begin{abstract}\noindent
A brief review is presented of the stability domain of three- and
four-charge ground-states when the constituent masses vary.
Rigorous results are presented, based on the scaling behavior and the
convexity properties deduced from the variational principle. They are 
supplemented by accurate numerical computations.
\end{abstract}
\vfill\vfill
\footnoterule
\vspace*{-.15cm}
\begin{center}
\begin{minipage}{.90\textwidth}
\setlength{\baselineskip}{6pt}
\footnotesize\noindent
${}^{*}$Talk
 given at the Workshop {\em Critical Stability}, Les Houches, Oct.\
 8--12, 2001, to appear in the Proceedings, a special issue of {\em
 Few-Body Systems}\\[2pt]
${}^\dagger${{\em E-mail:} Jean-Marc.Richard@isn.in2p3.fr}\\[2pt]
${}^\ddagger${Permanent address}
\end{minipage}\end{center}
\end{titlepage}
\section{Three unit-charge systems}
\label{three-unit}
Beyond hydrogenic atoms, the simplest systems in molecular physics
consist of three unit charges, $\pm(+1,\,-1,\,-1)$. The Hamiltonian
reads
\begin{equation}
\label{H-3-u}
H={\vec{p}_1^2\over 2 m_1}+{\vec{p}_2^2\over 2 m_2}+
{\vec{p}_3^2\over 2 m_3}-{e^2\over r_{12}}-{e^2\over r_{13}}+
{e^2\over r_{23}}~\cdot
\end{equation}
A non-relativistic point-charge interaction is assumed, and everything
else is neglected: relativistic corrections, finite-size effects,
annihilation, strong interaction, etc.

Nevertheless, as pointed out by Thirring \cite{Thirring79}, it is not
completely obvious why some systems are stable,
e.g., $\mathrm{H}_2^+=(p,p,e^-)$, $\mathrm{Ps}^-=(e^+,e^-,e^-)$ or
$\mathrm{H}^-=(p,e^-,e^-)$, while others, such as $(p,\bar{p},e^-)$ or
$(e^-, p, e^+)$, immediately dissociate into a neutral atom and an
isolated charge.

One can use the scaling properties of (\ref{H-3-u}) and similar
Hamiltonians to choose $\hbar=e^2=1$, as well as $\sum\alpha_i=1$,
where $\alpha_i=1/m_i$ is the inverse mass of particle $i$.  It is
natural to use inverse masses, as the Hamiltonian depends linearly
upon them.  The condition $\sum \alpha_i=1$ is reminiscent of energy
conservation in three-body decays, any possible configuration of which
can be represented in a ``Dalitz plot'', which is an equilateral
triangle of unit height.  Each inverse mass is read as the distance to
a side, as shown in Fig.~\ref{trian1}.
\begin{figure}[!htpc]
 \begin{minipage}{0.45\textwidth}
   \begin{center}
      \setlength{\unitlength}{0.20pt}
          \begin{picture}(900,700)(200,50)
          \font\gnuplot=cmr10 at 10pt\gnuplot
          \put(300,100){\line(1,0){600}}
          \put(300,100){\line(3,5){300}}
          \put(900,100){\line(-3,5){300}}
          \put(590,630){\makebox(0,0)[l]{A$_1$}}
          \put(910,70){\makebox(0,0)[l]{A$_2$}}
          \put(270,70){\makebox(0,0)[l]{A$_3$}}
          \put(700,200){\line(0,-1){100}}
          \put(700,200){\line(-5,3){250}}
          \put(700,200){\line(5,3){103}}
          \put(710,150){\makebox(0,0)[l]{$\alpha_1$}}
          \put(550,250){\makebox(0,0)[l]{$\alpha_2$}}
          \put(700,250){\makebox(0,0)[l]{$\alpha_3$}}
          \end{picture}
          \caption{\label{trian1} 
         The domain of normalized inverse masses $\alpha_i$.}
    \end{center}
\end{minipage}
\hspace{1cm}
\begin{minipage}{.45\textwidth}
     \begin{center}
       \setlength{\unitlength}{0.20pt}
        \begin{picture}(900,700)(200,50)
        \font\gnuplot=cmr10 at 10pt\gnuplot
  \linethickness{.7pt}
  \put(300,100){\line(1,0){600}}
  \put(300,100){\line(3,5){300}}
  \put(900,100){\line(-3,5){300}}
   \linethickness{.4pt}
  \qbezier[120](600,100)(600,400)(600,600)
  \put(590,630){\makebox(0,0)[l]{A$_1$}}
  \put(910,70){\makebox(0,0)[l]{A$_2$}}
  \put(270,70){\makebox(0,0)[l]{A$_3$}}
   \linethickness{1.2pt}
   \qbezier[120](510,450)(600,250)(580,100)
   \qbezier[120](690,450)(600,250)(620,100)
       \end{picture}
\caption{\label{trian2} 
Schematic shape of the stability domain.}
\end{center}
\end{minipage}
\end{figure}

Initially, some incorrect results where obtained for these systems,
though these were sometimes based on very astute reasoning.  It is,
indeed, a more delicate operation to estimate numerically where the
stability frontier is than to calculate the binding energy of a system
whose stability is well established.  Recent numerical calculations
are, however, very accurate and reliable.  See, e.g,
Refs.~\cite{Korobov00,Frolov86,Frolov99} for some examples and the
results in the literature, and Ref.~\cite{Armour93} for a
comprehensive survey.  If one summarizes the results of the
literature, one finds that the stable configurations belong to a band
around the symmetry axis $m_2=m_3$ where like-sign charges bear the
same mass.

The shape of the stability domain, as schematically pictured in
Fig.~\ref{trian2}, results from three basic properties
\cite{Martin92}:
\begin{enumerate}
\item 
Any state along the symmetry axis is stable. This was shown by
  Hill \cite{Hill77} using the trial wave function
\begin{equation}\label{Hill-w-f}
\Psi=\exp\left(-a r_{12}- b r_{13}\right) + 
\{ a \leftrightarrow b\}~,
\end{equation}
where the possibility of having $a\neq b$ is crucial, as already noted
by Chandrasekhar \cite{Chandrasekhar44} for H$^-$ and Hylleraas
\cite{Hylleraas47} for Ps$^-$.  Note that if a wave function with only
an adjustable effective charge $Z_\mathrm{e}$, i.e., $\Psi\propto
\exp(-Z_{\mathrm e}(r_{12}+r_{13}))$, or even more generally, any
factorized wave function $f(r_{12}) f(r_{13})$ is used, it is not
possible to demonstrate the stability of H$^-$ \cite{ten-Hoor95}; one
needs the anticorrelation $a\neq b$, or an explicit $r_{23}$
dependence, or both.
\item Each instability region is convex. This results from the
  convexity properties of the ground-state energy, after suitable
  rescaling such that the threshold energy $E(m_1^+,m_2^-)$ (or
  $E(m_1^+,m_3^-)$ on the other side) is kept constant.
  
\item Each instability region has as a  star shape with respect to A$_2$
  or A$_3$. For instance, if one draws a line from A$_3$ toward the
  symmetry axis, it corresponds, after rescaling, to increasing $m_3$
  while the threshold energy $E(m_1^+,m_2^-)$ remains constant.
  Increasing $m_3$ decreases the three-body binding energy and thus
  improves stability.
\end{enumerate}

\begin{figure}[!!!!tpc]
\begin{minipage}{.41\textwidth}
\begin{center}
 \setlength{\unitlength}{0.30pt}
  \begin{picture}(700,700)(150,0)
  \font\gnuplot=cmr10 at 10pt
  \gnuplot
  \linethickness{.7pt}
  \put(300,100){\line(1,0){400}}
  \put(300,100){\line(3,5){300}}
   \linethickness{.4pt}
  \qbezier[120](600,100)(600,400)(600,600)
  \qbezier[40](450,350)(495,325)(675,225)
  \qbezier[40](510,450)(545,275)(580,100)
  \put(590,630){\makebox(0,0)[l]{A$_1$}}
   \put(270,70){\makebox(0,0)[l]{A$_3$}}
\linethickness{1.2pt}
 \qbezier[120](510,450)(600,250)(580,100)
\put(560,70){\makebox(0,0)[l]{$\beta$}}
\put(475,455){\makebox(0,0)[l]{$\alpha$}}
\put(510,280){\makebox(0,0)[l]{$\gamma'$}}
\put(570,300){\makebox(0,0)[l]{$\gamma$}}
\end{picture}
\end{center}
  \vskip .3cm
\caption{\label{trian3} 
Some critical points of interest.}
\end{minipage}
\hspace{1cm}
\begin{minipage}{.41\textwidth}
\begin{center}
\includegraphics*[width=1.1\textwidth]{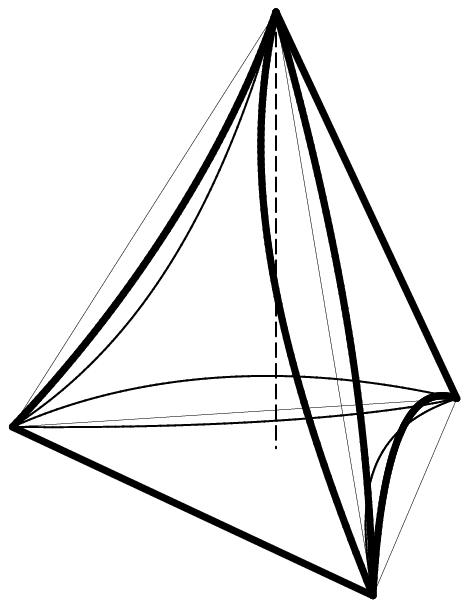}
\caption{\label{tetra1}Schematic shape of the stability domain of 
  $(m_1^+,m_2^+,m_3^-,m_4^-)$ states in the tetrahedron of normalized
  inverse masses $\alpha_i=m_i^{-1}$, with $\sum\alpha_i=1$.}
\end{center}
\end{minipage}
\end{figure}

An application of this approach can be made when searching the
critical masses for which stability disappears, for instance,
\begin{equation}
\alpha:\  (m_\alpha^+,\infty^-,1^-)\quad
\beta:\ (\infty^+,m_\beta^-,1^-)\quad
\gamma:\ (1^+,m_\gamma^-,1^-)~.
\end{equation}
In Fig.~\ref{trian3}, one can see that a safe (lower) bound on
$m_\alpha$ and a safe (upper) bound on $m_\beta$ result in a straight
line entirely in an instability domain.  It intersects at $\gamma'$
the $m_1=m_3$ axis, this providing a upper bound for the critical mass
$m_\gamma$, which turns out to be better \cite{Martin92} than the
limits obtained by direct studies of the particular configurations
$(1^+,m^-,1^-)$.  In other words, global considerations on the
triangle of stability are sometimes more powerful than local studies.


\section{Four unit-charge systems}
\label{four-unit}
Consider now $(m_1^+,m_2^+,m_3^-,m_4^-)$ systems, whose
Hamiltonian reads
\begin{equation}
\label{H-4-u}
H={\vec{p}_1^2\over 2 m_1}+{\vec{p}_2^2\over 2 m_2}+
{\vec{p}_3^2\over 2 m_3}+{\vec{p}_4^2\over 2 m_4}
-{e^2\over r_{13}}-{e^2\over r_{23}}-{e^2\over r_{14}}-{e^2\over 
r_{24}}+
{e^2\over r_{12}}+{e^2\over r_{34}}~\raise 2pt\hbox{,}
\end{equation}
with, again, some scaling properties, so that $\sum\alpha_i=1$ can
be imposed: each configuration is a point in a regular tetrahedron,
and the inverse constituent masses $\alpha_i=m_i^{-1}$ represent the
distances to the faces.

The lowest threshold consists of either two atoms or one three-body
ion and an isolated charge. The lowest two-atom state is obtained by
combining the heaviest of the positive charges $m_1$ and $m_2$ with
the heaviest of the negative charges $m_3$ and $m_4$, and allowing the
two lightest masses to form a second atom. This can be seen explicitly
from the Bohr formula, but this is a more general results for any
given potential, as shown in Refs.~\cite{Bertlmann80,Nussinov84}. If
for instance the first three particles are heavy and form a stable
ion, then the lowest threshold is $(m_1^+,m_2^-,m_3^-)+m_4^-$. As the
ion attracts the last charge with a potential that is asymptotically
attractive and Coulombic, and thus supports many bound states, one is
sure in such situation that the four-body system is stable. In
practice, stability is guaranteed by finding a wave function whose
expectation value for $H$ is smaller than the energy of the lowest
two-atom threshold.

For the equal-mass case, i.e., the positronium molecule Ps$_2$, this
was done in 1947 by Hylleraas and Ore \cite{Hylleraas47a}.
Later on, it was realized that the stability of Ps$_2$ implies that of
all hydrogen-like configurations $(M^+,M^+,m^-,m^-)$, $\forall M/m$.
This includes the ordinary hydrogen molecule, starting from $M=m$,
instead of the usual (and more natural) $M\to\infty$ limit of the
Born--Oppenheimer--Heitler--London treatment. The reasoning, made
independently by Adamowski et al.\ \cite{Adamowski71} and in
Ref.~\cite{Richard93}, is reminiscent of the well-known argument 
that
if an even Hamiltonian, say $h_\mathrm{even}=p^2+x^2$, is supplemented
by an odd term $h_\mathrm{odd}$, then the ground state is lowered. 
This
is seen explicitly if $h_\mathrm{odd}=x$, but the general result
 follows simply from the variational principle applied to
$h_\mathrm{even}+h_\mathrm{odd}$ using the even ground-state of
$h_\mathrm{even}$ as trial wave function.  Here, parity is replaced by
charge conjugation, and the $(M^+,M^+,m^-,m^-)$ Hamiltonian is
rewritten as
\begin{eqnarray}
&&H=H_\mathrm{even}+H_\mathrm{odd}=
\left[\left({1\over 4M}+{1\over 4m}\right)
(\vec{p}_1^2+\vec{p}_2^2+\vec{p}_3^2+\vec{p}_4^2)+V\right]\nonumber\\
&&\phantom{H=H_\mathrm{even}+H_\mathrm{odd}}
{}+\left[\left({1\over 4M}-{1\over 4m}\right)
(\vec{p}_1^2+\vec{p}_2^2-\vec{p}_3^2-\vec{p}_4^2)\right]~.
\end{eqnarray}

The full Hamiltonian $H$ has a lower ground-state than its even part
$H_\mathrm{even}$, which is nothing but a rescaled version of the
Ps$_2$ case. The nice feature is that $H$ and $H_\mathrm{even}$ have
the same threshold, since when one computes the two-body energy, one
first averages the inverse masses $M^{-1}$ and $m^{-1}$ to calculate
the reduced mass.

Once the binding energy of $(M^+,M^+,m^-,m^-)$ is given, then minimal
extensions of the stability domain beyond the line $\{ m_1=m_2,\,
m_3=m_4\}$ can be derived \cite{Richard94,Bressanini98}.

Many other configurations have been studied in the literature.  In
particular, all those with $m_3=m_4$ have been found to be stable. 
This property was checked extensively in Ref.~\cite{Varga97}.  Thus,
for two identical electrons ($m_3=m_4)$, the molecule is always
stable, irrespective of whether are the masses of the nuclei are equal
or unequal, heavier or lighter than the electrons, or even one lighter
and one heavier.

The results obtained can be very conveniently described using the 
tetrahedron of Fig.~\ref{tetra1}, which is also referred to
by Armour \cite{Armour01}. The thick lines correspond to stable configurations
with $\alpha_1=\alpha_2=0$ or $\alpha_3=\alpha_4=0$. The middle is
just the familiar hydrogen molecule with infinitely heavy protons. The
thin sides are outside the stability domain. The configuration at the
middle of a thin line corresponds to $(p,e^+,\bar{p},e^-)$, studied in
detail in \cite{Armour01}. The protonium $(p\bar{p})$ forms a neutral
and compact object with much too small an internuclear separation to 
bind  the
positronium atom $(e^+e^-)$.

\section{Further studies}
\label{further}
So far, we have restricted ourselves to the ground state and to unit
charge. Several generalizations can be envisaged.

The case of three particles with arbitrary charges
$\pm(q_1,-q_2,-q_3)$ is discussed rather extensively 
in Refs.~\cite{Martin95,Krikeb00}.
  
  In principle, our study can be repeated for excited states. For the
  first state with specific quantum numbers (negative parity, or a
  given value of the total angular momentum), the convexity properties
  remain, as one is dealing with the ground state in another sector 
of the
  Hilbert space. For the second or third state of given quantum
  numbers, the situation is more delicate.
  
  For excited states, stability should be addressed with respect to 
the
  appropriate threshold. For instance, a three-charge state of {\em
    unnatural} parity  $J^P=1^+$ cannot decay into the
  ground state of an atom and an isolated charge, unless a photon is
  also emitted. If one neglects radiative processes, then the lowest
  threshold involves an excited atom and is thus much higher than the
  threshold relevant for natural-parity states.
  
  The rigorous results on these systems are rather limited. There
  are fortunately more and more powerful methods which can be used for
  numerical investigations, among which are the stochastic variational
  method \cite{Varga95} and Monte-Carlo methods \cite{Bressanini01}.
  Among the remarkable results obtained by sophisticated numerical
  methods, we already mentioned the evidence that
  $(m_1^+,m_2^+,1^-,1^-)$ systems are stable $\forall\,  m_1,\,m_2$.
  There is also the discovery of stable excited states for the
  Positronium molecule \cite{Varga98}.
  
  Coulomb systems with more than four particles have been
  investigated. In the case of bosons, say $(m^+,m^-)^N$, it has been known
  for long-time that the system becomes more and more bound when the
  number $2N$ of constituent increases. A kind of precocious approach
  to the large $N$ limit was found in Ref.~\cite{Fleck95}. More
  precise binding energies are provided in Ref.~\cite{Varga95}. In
  more recent papers, Varga et al.\ revealed several new stable
  configurations with an atom or an ion binding two positrons
  \cite{Varga99,Mezei01}.
  
  The methods can also be applied to other potentials. For power-law
  potentials,  the scaling laws are comparable to those of the 
Coulomb case,
  and the stability region can be drawn  in a triangle for three-body
  systems, and a tetrahedron for four bodies, with similar convexity
  properties. Otherwise, the representation acquires additional
  dimensions. 

  The role of the number of space dimensions has also been studied 
\cite{Varga98a}.

\vskip .5cm{\small
    I would like to thank A.~Martin, Tai T.~Wu, J.~Fr{\"o}hlich,
    G.M.~Graf, M.~Seifert A.~Krikeb, S.~Fleck and K.~Varga for their
    enjoyable collaboration on this subject, and A.E.G.~armour and
    A.S.~Jensen for useful comments on the manuscript.
}    

\let\section=\oldsection
\renewcommand{\refname}

{\small References}
\begin{thebibliography}{10}
\itemsep-4pt
\footnotesize
\bibitem{Thirring79}
W.~Thirring, {\it A Course in Mathematical Physics}, {\bf Vol.~3}: Quantum
  Mechanics of Atoms and Molecules (Springer Verlag, New-York, 1979).

\bibitem{Korobov00}
V.I. Korobov, Phys. Rev. A {\bf 61} (2000) 064503.

\bibitem{Frolov86}
A.M. Frolov, Z. Phys. {\bf D2} (1986) 61.

\bibitem{Frolov99}
A.M. Frolov, Phys. Rev. A {\bf 59} (1999) 4270.

\bibitem{Armour93}
E.A.G. Armour and W. Byers Brown, Accounts of Chemical Research, {\bf 26}
  (1993) 168.

\bibitem{Martin92}
A.~Martin, J.-M.~Richard and T.T.~Wu, Phys.\ Rev.\ {\bf A46} (1992) 3697.

\bibitem{Hill77}
R.N.~Hill, J.~Math.~Phys.\ {\bf 18} (1977) 2316.

\bibitem{Chandrasekhar44}
S. Chandrasekhar, Astr. J. {\bf 100} (1944) 176; see, also, S. Chandrasekhar,
  {\it Selected papers}, vol.2 (The University of Chicago Press, Chicago,
  1989).

\bibitem{Hylleraas47}
E.A.~Hylleraas, Phys.\ Rev.\ {\bf 71} (1947) 491.

\bibitem{ten-Hoor95}
M. J. ten Hoor, Am. J. Physics, {\bf 63} (1995) 647.

\bibitem{Bertlmann80}
R. Bertlmann and A. Martin, Nucl.~Phys.~{\bf B168} (1980) 11.

\bibitem{Nussinov84}
S. Nussinov, Phys. Rev. Lett. {\bf 52} (1984) 966.

\bibitem{Hylleraas47a}
E.A.~Hylleraas and A.~Ore, Phys.\ Rev.\ {\bf 71} (1947) 493.

\bibitem{Adamowski71}
J. Adamowski, S. Bednarek and M. Suffczy{\'n}ski, Solid State Commun. {\bf 9}
  (1971) 2037; Phil. Mag. {\bf 26} (1972) 143.

\bibitem{Richard93}
J.-M. Richard, J.~Fr{\"o}hlich, G.M. Graf, and M. Seifert, Phys. Rev. Lett.
  {\bf 71} (1993) 1332.

\bibitem{Richard94}
J.-M. Richard, Phys. Rev. {\bf A49} (1994) 3573.

\bibitem{Bressanini98}
D. Bressanini, M. Mella and G. Morosi, Phys. Rev. A {\bf 57} (1998) 4956.

\bibitem{Varga97}
K. Varga, S. Fleck and J.-M. Richard, Europhys. Lett. {\bf 37} (1997) 183.

\bibitem{Armour01}
E.A.G. Armour, Contribution to this Workshop.

\bibitem{Martin95}
A.~Martin, J.-M.~Richard and T.T.~Wu, Phys. Rev. A {\bf 52} (1995) 2557.

\bibitem{Krikeb00}
A.~Krikeb, A.~Martin, J.-M.~Richard and T.T.~Wu, Few-Body Systems {\bf 29}
  (2000) 237.

\bibitem{Varga95}
K. Varga and Y. Suzuki, Phys. Rev. {\bf C52} (1995) 2885.

\bibitem{Bressanini01}
D. Bressanini, Contribution to this Workshop.

\bibitem{Varga98}
K. Varga, J. Usukura and Y. Suzuki, Phys. Rev. Lett. {\bf 80} (1998) 1876.

\bibitem{Fleck95}
S. Fleck and J.-M. Richard, Few-Body Systems {\bf 19} (1995) 19.

\bibitem{Varga99}
K. Varga, Phys. Rev. Lett. {\bf 83} (1999) 5471.

\bibitem{Mezei01}
J.Zs. Mezei et al., Phys. Rev. A {\bf 64} (2001) 032501.

\bibitem{Varga98a}
K. Varga, Phys. Rev. B {\bf 57} (1998) 13305.

\end{thebibliography}
 \end{document}